\documentclass[twocolumn]{jpsj2}

\def\runauthor{
Kenji {\sc Ohoyama}, Kentaro {\sc Indoh}, Aya {\sc Tobo}, Koji {\sc Kaneko}, Aya {\sc Hino} and Hideya {\sc Onodera}
}

\title
{
Detailed Measurements of Characteristic Profiles of Magnetic Diffuse Scattering in ErB$_2$C$_2$
}
\author
{ 
Kenji {\sc Ohoyama}\thanks{E-mail:ohoyama@imr.edu}, Kentaro {\sc Indoh}\thanks{Present Address: Schlumberger K.K., Sagamihara, Kanagawa 229-0006}, Aya {\sc Tobo}$^1$, Koji {\sc Kaneko}$^2$, Aya {\sc Hino} \\ and Hideya {\sc Onodera}$^1$
}
\inst
{
Institute for Materials Research, Tohoku University, Sendai 980-8577\\
$^{1}$Department of Physics, Graduate School of Science, Tohoku University, Sendai, 980-8578\\
$^{2}$Advanced Science Research Center, Japan Atomic Energy Research Institute, Tokai, 319-1195\\
}

\recdate
{
\today
}

\abst
{
Detailed neutron diffraction measurements on a single crystalline ErB$_2$C$_2$ were performed.
We observed magnetic diffuse scattering which consists of three components just above the transition temperatures, which is also observed in characteristic antiferroquadrupolar ordering compounds HoB$_2$C$_2$ and TbB$_2$C$_2$.
The result of this experiments indicates that the antiferroquadrupolar interaction is not dominantly important as a origin of the magnetic diffuse scattering.
}

\kword
{
ErB$_2$C$_2$, TbB$_2$C$_2$, HoB$_2$C$_2$, antiferroquadrupolar ordering, diffuse scattering, long periodic magnetic structure, neutron diffraction
}

\begin{document}
\maketitle
\section{Introduction}\label{introduction}
In this paper, we report results of detailed neutron diffraction experiments for the rare earth compound ErB$_2$C$_2$, and discuss anomalous magnetic diffuse scattering which is also observed in the isostructural compounds, TbB$_2$C$_2$ and HoB$_2$C$_2$, which exhibit characteristic antiferroquadrupolar (AFQ) orderings.

The tetragonal compounds RB$_2$C$_2$ (R=rare earth) with $P4/mbm$ symmetry~\cite{rf:LaB2C2, rf:LaB2C2_Kaneko, rf:Crystal_ohoyama} show various magnetic and quadrupolar behaviour due to coexistent competition between AFQ and antiferromagnetic (AFM) interactions.
In particular, AFQ orderings in HoB$_2$C$_2$ and TbB$_2$C$_2$ are characteristic.
HoB$_2$C$_2$ exhibits an AFQ ordering at $T_{\rm Q}$=4.5\,K under zero magnetic field, even though a magnetic ordered state exists between $T_{\rm Q}$ and $T_{\rm N}$=5.9\,K($>T_{\rm Q}$)~\cite{rf:HoB2C2, rf:HoB2C2_Shimada, rf:Ho11B2C2_ND}, which means that the AFQ ordering is realised under the internal magnetic field caused by the magnetic ordering.
The AFQ ordering in  HoB$_2$C$_2$ was also confirmed by resonant X-ray scattering experiments by Matsumura \textit{et al.}~\cite{rf:X_Matsu_Ho}.
On the other hand, TbB$_2$C$_2$ exhibits an AFM ordering at $T_{\rm N}$=21.7\,K, but shows no evidence of an AFQ ordering under zero magnetic field so far~\cite{rf:TbB2C2_JPSJ}.
TbB$_2$C$_2$ is, however, the unique compound which shows a magnetic field induced AFQ ordering~\cite{rf:TbB2C2_PRB}.
Since internal and external magnetic fields lift the degeneracy of the ground states which are required for AFQ orderings, the AFQ orderings in TbB$_2$C$_2$ and HoB$_2$C$_2$ are notable phenomena to understand mechanism of AFQ orderings in rare earth compounds.

We place a special emphasis on magnetic long periodic states in TbB$_2$C$_2$ and HoB$_2$C$_2$, because the periodicity in the $c$-plane is nearly the same, and anomalous magnetic scattering components were observed by neutron diffraction experiments in both the compounds.
The magnetic ordered state in TbB$_2$C$_2$ below $T_{\rm N}$ has a long periodic magnetic component as an appendant of a much dominant ${\mib k}$=(1,0,1/2) type AFM structure with a small ${\mib k'}$=(0,0,1/2) component; the propagation vector of the long periodic component is ${\mib k_{\rm L}}$=(1+$\delta$,\,$\delta$,\,0), where $\delta$=0.13$\sim$1/8~\cite{rf:TbB2C2_JPSJ}.
Moreover, Kaneko \textit{et al.} found an anomalous magnetic diffuse component which exists only in the square region in the reciprocal space surrounded by the four satellite peaks which correspond to the long periodicity.~\cite{rf:TbB2C2_TAS1}.
The magnetic diffuse scattering consists of at least three components: (i) broad peaks centred at the satellite positions which is probably due to short range ordering, (ii) ridge-like component between the satellite positions, and (iii) nearly flat component centred at the (1,0,0) reciprocal position.
Note that the ridge-like and flat components remain even at about 50\,K$\sim$2$T_{\rm N}$, while the broad peaks centred at the satellite positions disappear about 22\,K near $T_{\rm N}$.

HoB$_2$C$_2$ also shows similar anomalous magnetic diffuse scattering in the long periodic intermediate phase between $T_{\rm N}$ and $T_{\rm Q}$; the propagation vector is ${\mib k}$=(1+$\delta$,\,$\delta$,\,$\delta'$), where $\delta$=0.112$\sim$1/9, which is nearly the same as $\delta$ for TbB$_2$C$_2$, and $\delta'$=0.04~\cite{rf:ASR_Tobo, rf:HoB2C2_6G, rf:Ho11B2C2_ND}.
The flat top structure around the (1,0,0) reciprocal position is more obvious in HoB$_2$C$_2$ than that in TbB$_2$C$_2$, while the ridge structure is relatively not obvious for HoB$_2$C$_2$~\cite{rf:ASR_Tobo, rf:HoB2C2_6G}.
%

Note that the profile of the anomalous magnetic diffuse scattering observed in TbB$_2$C$_2$ and HoB$_2$C$_2$, in particular, the profile of the ridge-like and flat components can not be understood by only common magnetic correlations, for instance, magnetic short range ordering, or critical scattering near $T_{\rm N}$; the origin of the diffuse scattering component is not clear at the moment.
We think that it is worthy to study the origin of the magnetic diffuse scattering with strange profiles in TbB$_2$C$_2$ and HoB$_2$C$_2$ because the close periodicity and similar profiles suggest that the magnetic diffuse scattering is a remarkable character of the RB$_2$C$_2$ system.

To understand properties in TbB$_2$C$_2$ and HoB$_2$C$_2$, we think that  ErB$_2$C$_2$, the target compound in this paper, is important.
ErB$_2$C$_2$ exhibits a magnetic transition at $T_{\rm N}$=15.9\,K and a successive magnetic order-order transition at $T_{\rm t}$ = 13.0\,K; however, no AFQ ordering is realised~\cite{rf:ErB2C2, rf:Sakai2, rf:Duijn}.
The magnetic structure in the intermediate phase between $T_{\rm N}$ and  $T_{\rm t}$ is a transverse sinusoidal modulation type one with the magnetic moment parallel to the $c$-axis; the propagation vector is ${\mib k}$=(1+$\delta$,\,$\delta$,\,0), where $\delta$=0.112~\cite{rf:ErB2C2,rf:Duijn}, which is nearly the same as $\delta$ in TbB$_2$C$_2$ and HoB$_2$C$_2$.
Therefore, the close periodicity in the $c$-plane means that the long periodic states in ErB$_2$C$_2$, TbB$_2$C$_2$ and HoB$_2$C$_2$ should be compared to understand the magnetism in the long periodic states, though the magnetic anisotropy are different: the magnetic moments are parallel to the $c$-axis for ErB$_2$C$_2$, while those lie in the $c$-plane for HoB$_2$C$_2$ and TbB$_2$C$_2$.

Moreover, in our previous neutron powder diffraction experiments in ErB$_2$C$_2$, magnetic diffuse scattering around the (1,0,0) reciprocal position was observed at 14\,K; the diffuse scattering remains even at $T$$\sim$2$T_{\rm N}$~\cite{rf:ErB2C2}.
The comparison among the three compounds, therefore, must clarify roles of AFM and AFQ interactions for anomalous magnetic diffuse scattering in TbB$_2$C$_2$ and HoB$_2$C$_2$, because no AFQ ordering is realised for ErB$_2$C$_2$ 

Thus, the purpose of this study is to observe detailed profiles of the diffuse scattering around the (1,0,0) reciprocal position in ErB$_2$C$_2$ by single crystal neutron diffraction experiments, and to discuss if AFQ interaction is important as a possible origin of the anomalous diffuse scattering observed in the AFQ ordering materials, TbB$_2$C$_2$ and HoB$_2$C$_2$.
\section{Experimental Details}\label{experiments}
For sample preparation, the mixtures of 99.9\% pure Er, 99.5\% enriched $^{11}$B and 99.999\% pure C were melted by the conventional argon arc technique; natural boron was replaced by enriched $^{11}$B isotope to avoid strong neutron absorption.
To make sure its homogeneity, each ingot was turned over and remelted several times. The single crystals were grown by the Czochralski method with a tri-arc furnace.

We performed neutron diffraction experiments on the two machine installed at the reactor, JRR-3M, in Japan Atomic Energy Research Institute, Tokai: the neutron diffractometer KSD constructed by Institute for Materials Research, Tohoku University, and the triple axis spectrometer TOPAN constructed by Faculty of Science, Tohoku University.
For the experiments on KSD, a neutron beam with $\lambda$=1.52\AA\, was obtained by the 3\,1\,1 reflection of the Ge monochromator, and the collimation condition was 12$'$-open-sample-30$'$.
For the experiments on TOPAN,  the data were obtained by the triple-axis mode for elastic condition to avoid contamination of inelastic scattering components.
A neutron beam of $E_i$=$E_f$=14.7\,meV were used by the pyrographite monochromator and analyser under a collimation condition of 30$'$-30$'$-Sample-30$'$-30$'$.
The single crystalline sample was sealed in an aluminium cell with He gas, and was mounted at the cold head of a closed cycle He-gas refrigerator as the $c$-plane to be horizontal.
\begin{figure}[t]
  \begin{center}
\includegraphics[width=8cm]{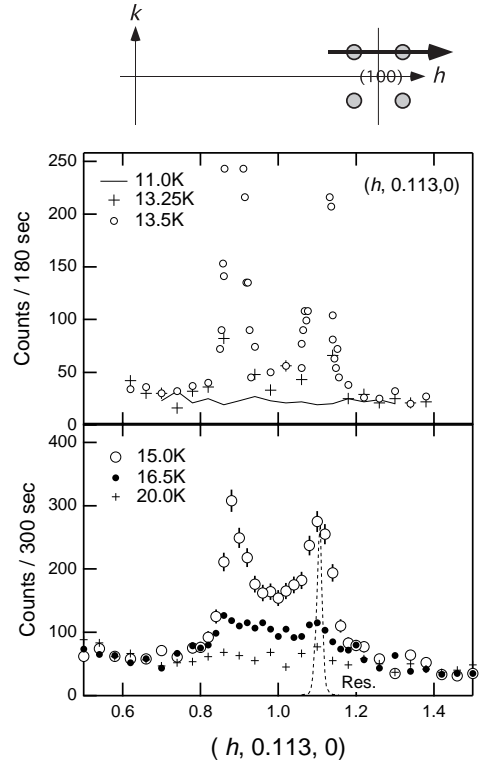}
  \end{center}
\caption{Magnetic scattering from a single crystalline sample of Er$^{11}$B$_2$C$_2$ around the (1,0,0) reciprocal position along the (1,0,0) direction through the satellite positions at several temperatures obtained on KSD.  The scan direction is shown above the figures with the arrow: the four gray circles  indicate the satellite positions.  The upper and lower figures indicate profiles observed below and above 15\,K.  The dashed line in the lower figure is the satellite peak at 13.5\,K (the same data in the upper figure) which corresponds to the resolution.}
\label{fig:fig1}
\end{figure}
\begin{figure}
  \begin{center}
	 \includegraphics[width=8cm]{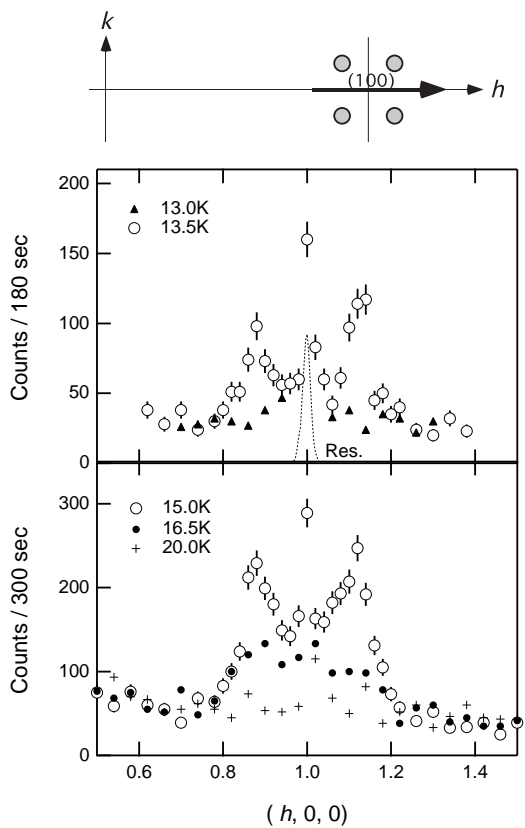}
  \end{center}
\caption{Magnetic scattering from a single crystalline sample of Er$^{11}$B$_2$C$_2$ along the (1,0,0) direction through the (1,0,0) reciprocal position obtained on KSD.  The upper and lower figures indicate profiles observed below and above 15\,K.  The dashed line in the upper figure is 100 magnetic Bragg peak at 2.2\,K which corresponds to the resolution.  A peak at the (1,0,0) observed at 15.0\,K is contamination of multiple nuclear scattering.}
\label{fig:fig2}
\end{figure}
\section{Results}\label{result}
Figures 1 and 2 show temperature dependence of profiles of the magnetic scattering along the (1,0,0) direction through the satellite positions at (1$\pm$0.11, 0.11,0), and through the (1,0,0) reciprocal position, respectively.
A point observed at the (1,0,0) reciprocal position in Fig.2 is contamination of multiple nuclear scattering, because it remains even above $T$=50\,K.
As shown in Fig.1, no obvious scattering was observed at $T$=11.0\,K.
With increasing temperature, the sharp satellite peaks which correspond to the long range and long periodic magnetic ordering rapidly develop above 13.0\,K.
We also observed development of diffuse scattering at the (1,0.113,0) reciprocal position, which is the middle point of the sharp satellite peaks.
At $T$=15.0\,K, the broader peaks than the resolution around the satellite positions, which can be distinguished from the sharp satellite peaks by its width, becomes obvious, though the sharp satellite peaks were no longer observed.
At $T$=16.5\,K, the magnetic diffuse scattering around (1,0.113,0) reciprocal position was still observed, though the broad peaks around the satellite positions were not clear.
Similar development of diffuse scattering was also observed measurements through the (1,0,0) reciprocal position as shown in Fig.2.
\begin{figure}
  \begin{center}
\includegraphics[width=7.5cm]{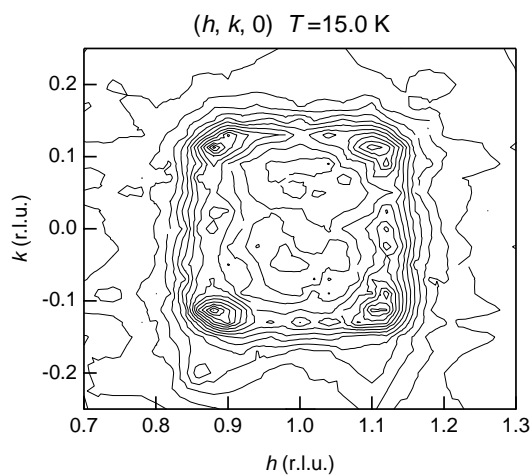}
  \end{center}
\caption{Contour map of the intensity of the magnetic scattering around the (1,0,0) reciprocal position in the ($h$, $k$, 0) plane at $T$=15.0\,K obtained on KSD.  The sharp satellite peaks due to the long range and long periodic ordering were not observed at this temperature.  The multiple scattering peak at the (1,0,0) reciprocal position was cut out from this contour plot.}
\label{fig:fig3}
\end{figure}

To show the general appearance of the diffuse scattering, distribution of the intensity around the (1,0,0) position in the ($h$, $k$, 0) plane is plotted in Fig.3.
The data were obtained at $T$=15.0\,K where the sharp satellite peaks due to the long range and long periodic ordering no longer exist.
As expected from the profiles in Figs.1 and 2, there exists intensity in the region surrounded by the satellite positions, and ridge-like component between two satellite positions, as well as the broad diffuse scattering around the satellite positions.
Note that this characteristic distribution of magnetic scattering is basically the same as those found in HoB$_2$C$_2$~\cite{rf:ASR_Tobo, rf:HoB2C2_6G}, and TbB$_2$C$_2$~\cite{rf:TbB2C2_TAS1}.

To discuss the profiles in more detail, we measured data along the (1,1,0) and (0,1,0) directions, because the conditions give much better resolution than that in the data shown in Figs.1 and 2.
Figures 4 show the profiles at (a) ($h$, $h$-1, 0), (b) (0.89, $k$, 0) and (c) (1.02, $k$, 0) at $T$=15.2\,K.
The scan directions are shown as the arrow in the insets; the meaning is the same as in Figs.1 and 2.
The small peak at (1,0,0) in Fig.4(a) is contamination due to multiple scattering.
As the resolution function for these experiments, the satellite peaks at (1.11, 0.11, 0) and at (0.89, 0.11, 0) at $T$=13.5\,K, and the 100 magnetic Bragg peak at $T$=5.9\,K are plotted in the figures as well.
As shown in the figures, broadening of the magnetic diffuse scattering obviously exceed the resolution limit; therefore, the resolution effect is negligible for these experiments.

In Fig.4(a), note that the profiles around the satellite positions at (0.89, -0.11, 0) and (1.11, 0.11, 0) are anisotropic; the slopes of the profiles in the region of $h$ $\ge$ 1.11 and $h \le$0.89 (outside region) are steep in comparison with those in the region of 0.89 $\le$ $h$ $\le $1.11 (inside region).
Moreover, there exists some structure in the inside region in Fig.4(b) and (c), particularly around $k$=0.05, which are not obvious in the contour map in Fig.3.
These characters of the profiles were also observed in experiments on other diffractometers.
\begin{figure}
  \begin{center}
\includegraphics[width=7.5cm]{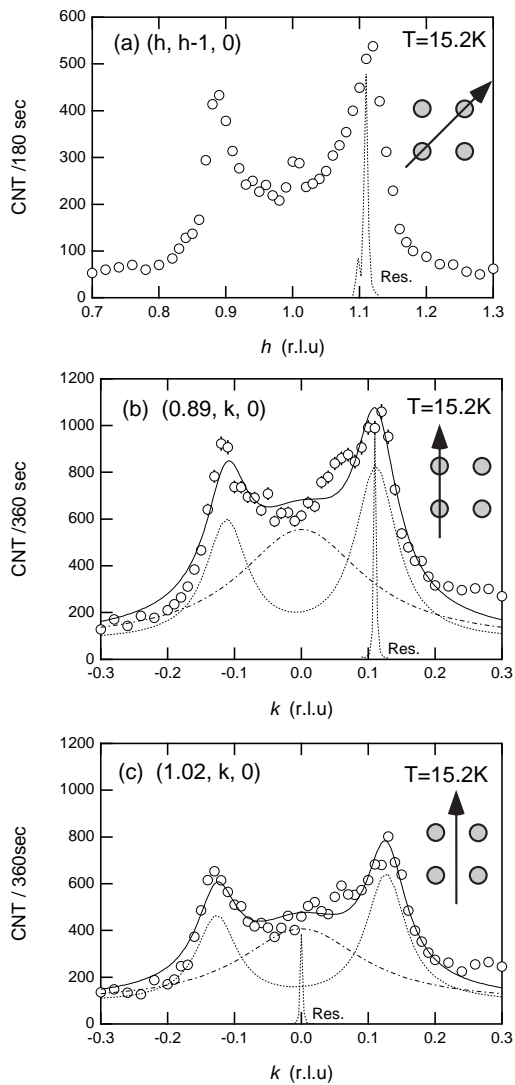}
  \end{center}
\caption{The profiles of the magnetic diffuse scattering at $T$=15.2K obtained on KSD (a) along the (1,1,0) direction through the satellite positions and (b) along the (0,1,0) direction through the satellite positions, and (c) along the (0,1,0) direction through the (1,0,0) reciprocal lattice position.  The scan directions are shown in the insets with the arrows: the four gray circles  indicate the satellite positions as Figs.1 and 2.  The 100 magnetic Bragg peak at $T$=5.9\,K and satellite peak at $T$=13.5\,K are also plotted as the resolution functions.  The solid lines indicate the best results of Lorentzian fitting; each Lorentzian components are plotted with the dotted and chained lines.}
\label{fig:fig_kscan}
\end{figure}

To confirm existence of the ridge-like component and the component centred at (1,0,0), we distinguished each component by least-square fitting.
The solid lines in Figs.4(b)(c) indicate the best results of the least-square fitting procedure with three Lorentzian functions: in Fig.4(b), two Lorentzians with the same width around (0.89, $\pm$0.11, 0) and a Lorentzian at (0.89, 0, 0), and in Fig.4(c), two Lorentzians with the same width around (1, $\pm$0.11, 0) and a Lorentzian at (1,0,0); each Lorentzian component is also shown in the figures with dashed and chained lines.

Although the fitting function in Figs.4(b) and (c) does not give complete agreement in the inside region because of the small structures around $k$=$\pm$0.05, we can conclude  at least following points.
Since sum of the two broad peaks centred at the satellite positions (dotted lines) can not represent the intensity of the inside region, the broad components around (0.89, 0,0) in Fig.4(b) and around (1,0,0) in Fig.4(c) are needed to describe the scattering in the inside region.
Thus, existence of the broad component around (1,0,0) and the ridge-like component between the satellite positions are undoubted.
These results of the fitting procedure and the distribution in Fig.3, therefore, indicate that the magnetic diffuse scattering in the ($h$, $k$, 0) plane consists of at least three components: (i) the broad peaks at the satellite positions, (ii) the ridge-like one around (1$\pm$0.11, 0, 0) and (1, $\pm$0.11, 0) reciprocal lattice position, and (iii) the broadest one centred at the (1,0,0) reciprocal lattice position.
Such profiles which consists of three broad components were also observed in the AFQ ordering compound TbB$_2$C$_2$~\cite{rf:TbB2C2_TAS1}, and HoB$_2$C$_2$~\cite{rf:HoB2C2_6G}.
 At the moment, we can not find any sufficient profile function which has clear physical meanings to represent the profiles in ErB$_2$C$_2$, TbB$_2$C$_2$ and HoB$_2$C$_2$.

Note again that the fitting function in Figs.4(b) and (c) can not represent the profiles completely, because of the small structures around $k$=$\pm$0.05.
We should point out that similar disagreement was also reported for TbB$_2$C$_2$~\cite{rf:TbB2C2_TAS1}; the magnetic diffuse scattering observed in TbB$_2$C$_2$ shows small shoulders at the foot of the broad peak in the inside region which can not be represented with simple profile functions (Figs.1(a) and (b) in ref. 10).

From the fitting procedures in Figs.4, the half width of each Lorentzian function was estimated to be (i) 0.038(8) r.l.u. for the broad peaks at the satellite positions, (ii) for the ridge-like one, 0.11(7) r.l.u. along the ridge line and 0.036(7) r.l.u. perpendicular to the ridge line, and (iii) 0.11(6) r.l.u. for the broadest one centred at the (1,0,0) reciprocal lattice position, when the resolution effect is neglected.
The errors were determined as the standard deviation calculated by the least square fitting procedure.
Assuming that the broad peaks at the satellite positions are due to short range correlation, we roughly estimated the correlation length as $\sim$20\AA, which is about four times larger than the lattice constant, $a$.
\begin{figure}
  \begin{center}
\includegraphics[width=7.5cm]{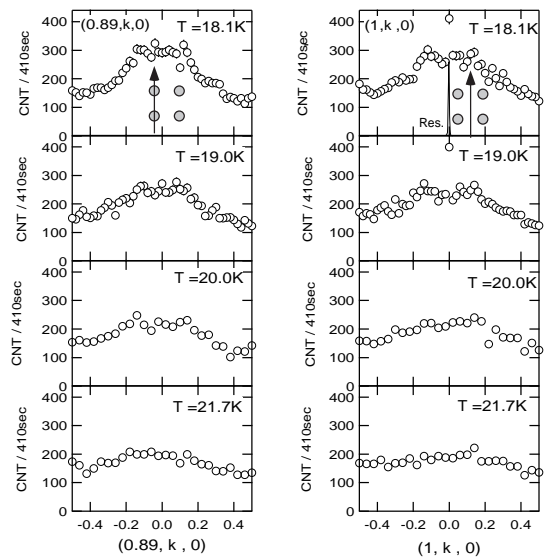}
  \end{center}
\caption{Temperature dependence of the profile of the magnetic diffuse scattering above 18\,K observed on TOPAN. The solid line in the right-top figure indicates 100 magnetic Bragg reflection, which correspond to the resolution.}
\label{fig:fig5}
\end{figure}

To confirm the profiles at higher temperature, we measured the magnetic diffuse scattering at higher temperatures on TOPAN, because the beam flux of TOPAN is much intense in comparison with that of KSD.
Figure~\ref{fig:fig5} show the profiles of the magnetic diffuse scattering at several temperatures above 18\,K observed on TOPAN.
At $T$=18\,K, the broad peaks which were observed around the satellite positions at 15\,K were not observed; consequently, the characteristic flat-top profile is obvious.
With increasing temperature, the diffuse scattering becomes weak, but the flat-top profile is still observed at 20\,K, while it is not obvious if the diffuse scattering remains at $T$=21.7\,K.
Thus, the magnetic diffuse scattering retains the flat-top profile up to about 20\,K, and disappears at a temperature near 22\,K $\sim$ 1.4$T_{\rm N}$.
Note that the magnetic diffuse scattering components for TbB$_2$C$_2$ and HoB$_2$C$_2$ also remain up to $\sim$2$T_{\rm N}$ or 2$T_{\rm Q}$~\cite{rf:TbB2C2_TAS1, rf:ASR_Tobo, rf:HoB2C2_6G}.

\section{Discussion}\label{discussion}
Let us discuss the origin of the magnetic diffuse scattering observed  in RB$_2$C$_2$.
We think that the diffuse scattering observed in the present study in ErB$_2$C$_2$ is basically the same type one as those in the characteristic AFQ ordering compounds HoB$_2$C$_2$ and TbB$_2$C$_2$ because of the following similarities on its profiles~\cite{rf:TbB2C2_TAS1,rf:ASR_Tobo, rf:HoB2C2_6G}.
ErB$_2$C$_2$ shows the anomalous diffuse scattering far above the transition temperatures which consists of at least three diffuse components: (i) the broad peak around the satellite peaks positions, (ii) the ridge-like component along (1,0,0) and (0,1,0) directions between two satellite positions, and (iii) the broadest component with nearly flat top structure centred at the (1,0,0) position.
Note that TbB$_2$C$_2$ and HoB$_2$C$_2$ show such multi-component diffuse scattering which remains up to $\sim 2T_{\rm N}$ as well. 
Moreover, a trapezium-like profile with flat top is also observed in HoB$_2$C$_2$~\cite{rf:ASR_Tobo,rf:HoB2C2_6G}.
The profile of the diffuse scattering in ErB$_2$C$_2$ can not be represented with simple profile functions satisfactory, for instance sum of Lorentzian or Gaussian functions because of some structures in the inside region.
Quite similar structures in the inside region was also observed in TbB$_2$C$_2$~\cite{rf:TbB2C2_TAS1}.
For the similarities on the profiles, thus, we conclude that the magnetic diffuse scattering observed in ErB$_2$C$_2$ has the same origin as those observed in the AFQ ordering compounds TbB$_2$C$_2$ and HoB$_2$C$_2$.

Although the origin of the diffuse scattering in ErB$_2$C$_2$, especially the ridge-like and flat-top components, is not clear, on the analogy of HoB$_2$C$_2$, the broad peaks around the satellite positions observed in ErB$_2$C$_2$ above the transition temperatures are probably caused by magnetic short range correlations.
In HoB$_2$C$_2$, a broad hump of specific heat but no sharp peak were observed at $T_{\rm N}$ where the satellite peaks which correspond to the long periodic ordering develop~\cite{rf:HoB2C2_6G}.
Moreover, the satellite peaks in HoB$_2$C$_2$ are much broader than the resolution limit.
For these points the long periodic state below $T_{\rm N}$ in HoB$_2$C$_2$ proved to be due to short range correlation.
Thus, we think that the broad peaks around the satellite positions observed above $T_{\rm N}$ in ErB$_2$C$_2$ is also due to short range correlation which develops above $T_{\rm N}$.

Before this study, the characteristic diffuse scattering in TbB$_2$C$_2$ and HoB$_2$C$_2$, particularly, the flat top and the ridge-like component \textit{were} thought to be a noteworthy character of the AFQ ordering compounds.
The present results for ErB$_2$C$_2$, however, indicate that the AFQ ordering is not necessarily required as the origin of the magnetic diffuse scattering, because no AFQ ordering is realised in ErB$_2$C$_2$.
Though the origin is still an open question, we should consider magnetic interactions between $4f$ electrons and/or some common character in RB$_2$C$_2$.
\section*{Acknowledgements}
The authors would like to thank Mr. K. Nemoto and Dr. H. Hiraka of IMR, Tohoku University for their helpful assistance in the neutron scattering experiments.

\end{document}